\documentclass[aps,twocolumn,preprintnumbers,groupedaddress,nofootinbib,pra]{revtex4-2}
\pdfoutput=1

\usepackage[usenames,dvipsnames,table]{xcolor}
\usepackage{graphicx,amsmath,amssymb,amsthm,multirow,array,bm,bbm,esint}
\usepackage{enumitem}
\usepackage[mathscr]{eucal}
\usepackage[bbgreekl]{mathbbol}
\usepackage{epsf,amsfonts}
\usepackage{physics}
\usepackage{dsfont, mathtools}
\usepackage{slashed}
\usepackage{hyperref}
\usepackage{mathdots}
\usepackage{upgreek}
\usepackage{comment}

\usepackage[normalem]{ulem}

\hypersetup{
    pdfstartview={FitH},    
    pdftitle={TITLE},    
    pdfauthor={Felix Haehl},     
    colorlinks=true,       
    linkcolor=blue,          
    citecolor=red,        
    filecolor=magenta,      
    urlcolor=blue           
}

\usepackage{tikz}

\newcommand{\beq}{\begin{equation}}
\newcommand{\eeq}{\end{equation}}

\begin{document}

\title{Modular-invariant random matrix theory and AdS$_3$ wormholes}

\author{Jan Boruch}
\affiliation{Leinweber Institute for Theoretical Physics and Department of Physics, University of California, Berkeley, CA 94720, USA}
\author{Gabriele Di Ubaldo}
\affiliation{Leinweber Institute for Theoretical Physics and Department of Physics, University of California, Berkeley, CA 94720, USA,}
\affiliation{RIKEN iTHEMS Center for Interdisciplinary Theoretical and Mathematical Sciences,\\ 2-1 Hirosawa, Wako, Saitama 351-0198 Japan,}
\affiliation{Universit\'{e} Paris-Saclay, CNRS, CEA, Institut de Physique Th\'{e}orique, 91191, Gif-sur-Yvette, France}
\author{Felix M.\ Haehl}
\affiliation{School of Mathematical Sciences and STAG Research Centre, University of Southampton, SO17 1BJ, U.K.}
\author{Eric Perlmutter}
\affiliation{Universit\'{e} Paris-Saclay, CNRS, CEA, Institut de Physique Th\'{e}orique, 91191, Gif-sur-Yvette, France,}
\affiliation{Institut des Hautes \'{E}tudes Scientifiques, 91440, Bures-sur-Yvette, France}
\author{Moshe Rozali}
\affiliation{Department of Physics and Astronomy, University of British Columbia, Vancouver, V6T 1Z1, Canada}

\begin{abstract}
We develop a nonperturbative definition of RMT${}_2$: a generalization of random matrix theory that is compatible with the symmetries of two-dimensional conformal field theory. Given any random matrix ensemble, its $n$-point spectral correlations admit a prescribed modular-invariant lift to RMT${}_2$, which moreover reduce to the original random matrix correlators in a near-extremal limit. Central to the prescription is a presentation of random matrix theory in Mellin space, which lifts to two dimensions via the $\text{SL}(2,\mathbb{Z})$ spectral decomposition employed in previous work. As a demonstration we perform the explicit RMT${}_2$ lift of two-point correlations of the GUE Airy model. We propose that in AdS$_3$ pure gravity, semiclassical amplitudes for off-shell $n$-boundary torus wormholes with topology $\Sigma_{0,n} \times S^1$ are given by the RMT${}_2$ lift of JT gravity wormhole amplitudes. For the three-boundary case, we identify a gravity calculation which matches the RMT${}_2$ result.
\end{abstract}

\maketitle


\section{Introduction}
\label{sec:intro}

Holographic duality for theories of two-dimensional gravity provides strong evidence that random matrix universality constrains the spectrum of holographic quantum systems \cite{Cotler:2016fpe,Saad:2019lba,Saad:2018bqo,Cotler:2020ugk}. The importance of two-dimensional conformal field theories (CFTs), dual to quantum gravity in AdS$_3$, for constructing a more robust holographic dictionary for high-energy states motivates the search for an inherently two-dimensional CFT definition of random matrix universality. Conversely, one seeks a framework to uplift any given random matrix theory (RMT) to 2D CFT in a manner that respects modular and conformal invariance. Following \cite{DiUbaldo:2023qli}, we shall refer to such a framework as RMT${}_2$. 

Just as RMT universally quantifies the spectral statistics of chaotic many-body systems in view of the Bohigas–Giannoni–Schmit Conjecture \cite{BGS1,BGS2}, RMT$_2$ should likewise describe the spectra of irrational 2D CFTs in sufficiently high-energy regimes. From a holographic perspective, RMT${}_2$ is expected to quantify the spectral statistics of primary operators dual to black hole states above extremality, generalizing arguments given in \cite{Maxfield:2020ale,Cotler:2020ugk} and built upon in \cite{Belin:2023efa,Jafferis:2024jkb}.

The construction of RMT$_2$ was initiated in \cite{Haehl:2023tkr,DiUbaldo:2023qli,Haehl:2023xys,Haehl:2023mhf}, which focused on the 2D CFT avatar of a central hallmark of random matrix universality, namely, the ``linear ramp'' of the spectral form factor (SFF). In this Letter, we provide a more complete definition of RMT$_2$: a prescription to uplift $n$-point RMT correlators to modular-invariant objects, without restricting to any regime of times or energies, which moreover reduce to the original RMT correlators in an appropriate near-extremal limit. The two-point case furnishes a modular-invariant SFF preserving a linear ramp and plateau structure. Further details and elaboration will be given in \cite{toappear}. 

\section{RMT${}_2$}
\label{sec:rmt2}

We will now describe a two-step procedure, which starts with a given RMT and constructs a corresponding modular-invariant RMT${}_2$.
Consider the $n$-point spectral form factor (SFF) of a given matrix Hamiltonian,
\beq
K^{(n)}_\text{RMT}(\{y_i\}) =  \left\langle \text{Tr}\left( e^{-y_1 H}\right) \cdots \text{Tr}\left(e^{-y_n H} \right)\right\rangle_\text{c}
\eeq
where $\langle \, \cdot \, \rangle_\text{c} \equiv Z^{-1} \int dH (\, \cdot \,) e^{-N \text{Tr}(H)}$ is the connected matrix integral and $y_i$ are inverse temperatures. The first step consists of computing the Mellin transform: 
\beq\label{eq:mellin}
\begin{split}
 f^{(n)}_\text{RMT}(\{\omega_i\}) &\equiv {\cal M}\left[ K_\text{RMT}^{(n)}(\{y_i\}) \right](-i\omega_1,\ldots,-i\omega_n)\,,
 \end{split}
\eeq
where ${\cal M}[K(y)](-i\omega) = \int_0^\infty dy \, y^{-1-i\omega} \, K(y)$. This quantity defines {\it RMT in Mellin space}.

The second step, the conceptual core of the prescription, is to regard the Mellin transform \eqref{eq:mellin} as a spectral overlap in a modular-invariant eigenbasis. 
The RMT${}_2$ spectral $n$-point function on $n$ tori with modular parameters $\tau_i := x_i + i y_i$, denoted ${\cal Z}^{(n)}_{\text{RMT}_2}(\{\tau_i\})$, is defined in general as follows:
\begin{equation}
\label{eq:ZRMT2defgen}
   {\cal Z}^{(n)}_{\text{RMT}_2}(\{\tau_i\})  = \sum_{{\bf m}} e^{-S_0({\bf m})}\,{\cal Z}^{(n)}_{\text{RMT}_2}(\{\tau_i\}|{\bf m})\,,
\end{equation}
As we explain further in the following paragraphs, its structure is straightforward: we enforce RMT statistics near extremality in every spin sector on all $n$ components (or ``boundaries''). These spin sectors are encoded in the label ${\bf m}:=(m_1,\ldots,m_n)\in\mathbb{Z}^n$, a set of integer spins, one per boundary. The constant $S_0({\bf m}) := \sum_i S_0(m_i)$ is a book-keeping parameter; one may regard $S_0(m_i)$ as near-extremal entropy of spin-$m_i$. 


The all-scalar component ($\bf{m}=\bf{0}$), on which we will focus in this Letter, is given as follows:
\begin{align}
\label{eq:ZRMT2def}
 {\cal Z}^{(n)}_{\text{RMT}_2}(\{\tau_i\}|{\bf 0}) &\equiv \!
  \int_{\mathbb{R}^n}\!\!\!\left(\prod_{i=1}^n \frac{d\omega_i}{(2\pi)}E_{\frac{1}{2}+i\omega_i}(\tau_{i})\right)_{\!\!\! \rm sym}  \!\!f^{(n)}_{\rm RMT}(\{\omega_i\}) \nonumber\\ &\quad + [\text{\tt zero modes}] + [\text{\tt cusp forms}] \,,
\end{align}
where ``sym" denotes symmetrization over $\{\tau_i\}$, and  
\beq
 E_s(\tau) = \sum_{j\geq 0} (2-\delta_{j,0}) \, \mathtt{a}_j^{(s)} \cos(2\pi j x) \sqrt{y} K_{s-\frac{1}{2}}(2\pi j y)
\eeq
are non-holomorphic Eisenstein series with spin-$j$ Fourier coefficients $\mathtt{a}_j^{(s)}$ left implicit.
On the critical line $s \equiv \frac{1}{2} + i\omega \in {\cal C}_\text{crit}$ with $\omega \in \mathbb{R}$, 
the Eisenstein series span the continuous part (``scattering states'') of the spectrum of the Laplacian on the fundamental domain ${\cal F} = \mathbb{H}/PSL(2,\mathbb{Z})$ with eigenvalues $s(1-s)$. 

The terms in \eqref{eq:ZRMT2def} denoted as $ [\text{\tt zero modes}]$ contain the part of the spectral decomposition that involves the trivial eigenfunction $\phi_0=\text{constant}$. Because the Eisenstein series has a pole at $s=1$ with constant residue $\text{vol}({\cal F})^{-1}$, spectral overlaps with the constant function may be obtained by taking iterated residues of the first line of \eqref{eq:ZRMT2def} at $\omega_i=1/2i$: for the $n$-point correlator, taking $m$ such residues generates terms with $n-m$ Eisenstein factors. For example, the case $n=2$ is 
\beq
\label{eq:ZRMT2def2}
\begin{split}
 & {\cal Z}^{(2)}_{\text{RMT}_2}(\{\tau_i\}|{\bf 0})= \frac{1}{\text{vol}({\cal F})^2} \; f^{(2)}_{\rm RMT}\left( \tfrac{1}{2i},\tfrac{1}{2i} \right)  \\
& + \frac{1}{\text{vol}({\cal F})}  \int_{\mathbb{R}} \frac{d\omega_1}{2\pi} \,   f^{(2)}_{\rm RMT}\left(\omega_1,\tfrac{1}{2i}\right) \left(E_{s_1}(\tau_1)+E_{s_1}(\tau_2)\right)\\
 & + \int_{\mathbb{R}^2} \! \! \frac{d\omega_1 d\omega_2}{(2\pi)^2} \,  f^{(2)}_{\rm RMT}(\omega_1,\omega_2) \left(E_{s_1}(\tau_1) E_{s_2}(\tau_2)\right)_{\rm sym} + \big[\substack{\text{\tt cusp}\\ \text{\tt forms}}\big]
\end{split}
\eeq
As we will see below, following other CFT contexts for the ``standard" case $n=1$ \cite{Collier:2022emf}, these zero mode terms will in fact be canceled in the genus expansion of RMT${}_2$.

Note that we can equivalently write the Eisenstein sector of \eqref{eq:ZRMT2def} as 
\beq
\label{eq:microcanonical}
\begin{aligned}
 {\cal Z}^{(2)}_{\text{RMT}_2}(\{\tau_i\}|\bf{0}) &\supset \left\langle \text{Tr}\left( e^{-H}|_{E(\tau_1)} \right)\cdots \text{Tr}\left( e^{-H}|_{E(\tau_n)} \right) \right\rangle_\text{c} \\
 e^{-H}|_{E(\tau)} &\equiv \int_{\mathbb{R}} \frac{d\omega}{2\pi} \, \Gamma(-i\omega) H^{i\omega} E_{s}(\tau) \,.
 \end{aligned}
\eeq
This is mathematically equivalent since ${\cal M}[e^{-y \lambda}](-i\omega) = \Gamma(-i\omega)\lambda^{i\omega}$ for every eigenvalue $\lambda\in \text{spec}(H)$, but this formulation makes it manifest that the Eisenstein part of RMT${}_2$ is in fact still a matrix integral.

The terms in \eqref{eq:ZRMT2def} denoted as $ [\text{\tt cusp forms}]$ refer to the contribution of Maass cusp forms $\phi_n(\tau)$, an infinite set of eigenfunctions which spans the discrete part of the eigen-spectrum (``bound states''). While Maass cusp forms are crucial for describing RMT statistics of the spinning spectrum \cite{Haehl:2023xys}, they have no scalar Fourier mode, and are subleading near extremality, as we now explain. \\

\paragraph*{\bf Near-extremal constraints.}
The above definition of RMT${}_2$ yields a modular-invariant ``lift" of RMT. Conversely, a crucial property of RMT${}_2$ is that the {\it near-extremal limit}  effectively reduces $E_{s_i}(\tau) \rightarrow \sqrt{y_i}\,y_i^{i\omega_i}$ in the integral \eqref{eq:ZRMT2def}: 
this turns \eqref{eq:ZRMT2def} into an inverse Mellin transform, and $ {\cal Z}^{(n)}_{\text{RMT}_2}(\{\tau_i\}|{\bf 0})$  reduces to the original $K_\text{RMT}(\{y_i\})$,
\beq
\label{eq:EisNE}
 {\cal Z}^{(n)}_{\text{RMT}_2}(\{\tau_i\}|{\bf 0}) \longrightarrow \sqrt{y_1 \dots y_n} \,K^{(n)}_\text{RMT}(\{y_i\})\,,
\eeq
where the factor $\sqrt{y_1\cdots y_n}$ arises because we consider modular-invariant CFT partition functions counting primary states only \cite{Benjamin:2021ygh}. The near-extremal limit (for any topology) turns out to be the same as in \cite{Ghosh:2019rcj}, where one rescales uniformly $y_i \rightarrow \gamma y_i$ and takes $\gamma \rightarrow \infty$.



The object ${\cal Z}^{(n)}_{\text{RMT}_2}(\{\tau_i\}|{\bf 0})$ is designed to implement RMT constraints near extremality in the scalar sector on all $n$ components. A full RMT$_2$ amplitude should entail a similar reduction to the same RMT input independently for every spin sector: this is precisely what the additional terms given in \eqref{eq:ZRMT2defgen} are encoding, where the multi-index ${\bf m}$ labels the spin sectors in which near-extremal RMT statistics are being enforced. In general, due to modular mixing -- that is, {\it away} from extremality -- the spin-${\bf m}$ terms contribute to spin-${\bf j\neq m}$ sectors of ${\cal Z}^{(n)}_{\text{RMT}_2}(\{\tau_i\}) $. 
This is visible from \eqref{eq:ZRMT2def}, where Eisenstein series and cusp forms have spinning components, but only the former survives in the near-extremal limit \eqref{eq:EisNE}. We give a further accounting of this, and the enforcement of RMT in spinning sectors, in \cite{toappear}. 

In this Letter, in which we focus on the Eisenstein sector of the ${\bf m=0}$ amplitude, we are therefore imposing RMT statistics near extremality in the ${\bf j=0}$ sector, via the modular-invariant lift \eqref{eq:ZRMT2def} of RMT.\\


\paragraph*{\bf Warmup: AdS$_3$ wormhole and linear ramp.}  
The simplest prediction of RMT is the leading approximation to the two-point function: 
\beq
\label{eq:Kramp}
 K^{(0,2)}_{\text{RMT}}(y_1,y_2) = \frac{\mathsf{C}_{\rm RMT}}{2\pi} \frac{\sqrt{y_1y_2}}{y_1+y_2} \,,
\eeq
where the constant $\mathsf{C}_{\rm RMT}$ encodes the RMT universality class (e.g., $\mathsf{C}_\text{GUE} = 1$, $\mathsf{C}_\text{GOE}=2$). This contains the linear ramp of the SFF at late Lorentzian times, plus an infinite set of corrections that resum to the full double-scaled RMT result \cite{Saad:2019lba}. 

In Mellin space,
\beq 
\label{eq:fCJ}
\begin{split}
 f^{(0,2)}_\text{RMT}(\omega_1,\omega_2) &= \frac{\mathsf{C}_{\rm RMT}}{2\cosh(\pi\omega_1)} \times  \pi\delta(\omega_1+\omega_2) \,.
\end{split}
\eeq
Via \eqref{eq:ZRMT2def2}, this defines the simplest universal contribution to  RMT${}_2$, a modular-invariant completion of \eqref{eq:Kramp}. 
To make this more explicit, we recall the result of \cite{Cotler:2020ugk}, where the $\mathbb{T}^2 \times I$ wormhole amplitude in AdS$_3$ pure gravity was found to be 
\beq 
\label{eq:CJresult}
 {\cal Z}^{(0,2)}_{\text{AdS}_3}(\tau_1,\tau_2) = \frac{\mathsf{C}_{\rm RMT}}{4\pi^2} \sum_{\gamma \in \text{SL}(2,\mathbb{Z})} \frac{\text{Im}(\tau_1) \text{Im}(\gamma \tau_2)}{|\tau_1+\gamma \tau_2|^2} \,.
\eeq
In $\text{SL}(2,\mathbb{Z})$ spectral space \cite{DiUbaldo:2023qli},  
\beq 
\label{eq:CJRMT2}
\begin{split}
{\cal Z}^{(0,2)}_{\text{AdS}_3}(\tau_1,\tau_2) &= \int_\mathbb{R}\frac{d\omega_1}{2\pi} \frac{d\omega_2}{2\pi} \, f^{(0,2)}_\text{RMT}(\omega_1,\omega_2) E_{s_1}(\tau_1) E_{s_2}(\tau_2) \\
&\quad\;\; + \sum_{n_1,n_2} f^{(0,2)}_\text{RMT}(\omega_{n_1}, \omega_{n_2})\, \phi_{n_1}(\tau_1) \phi_{n_2}(\tau_2)\,,
\end{split}
\eeq
where in the cusp form sector one replaces $\pi \delta(\omega_1 + \omega_2) \rightarrow \delta_{n_1,n_2}$. This takes precisely the form of an RMT${}_2$ amplitude. This example illustrates how the RMT${}_2$ formalism reveals essential features of the gravity amplitude \cite{DiUbaldo:2023qli,Haehl:2023mhf}, such as the encoding of the linear ramp in the simple condition $f^{(0,2)}_\text{RMT}(\omega_1,\omega_2) \sim \delta(\omega_1+\omega_2)\,e^{-\pi \omega_1}$ for large $|\omega_i|$, and the amplitude being the diagonal approximation (\`a la Berry \cite{berry}) to a CFT trace formula. In what follows we go ``beyond the ramp" by performing the RMT${}_2$ lift of full RMT correlators.

\section{Paradigmatic Example: Airy RMT$_2$} 
\label{sec:airy}

As a natural starting point for demonstration, we study the topological expansion for the simplest instance of RMT${}_2$: the lift of the Airy model in the GUE universality class. The RMT is defined by the spectral density
\beq
\rho(E) = \rho_0(E) e^{-S_0} \,,\quad \rho_0(E) = \frac{1}{2\pi}\sqrt{E}\,,
\eeq
where $S_0$ is a large parameter. The associated two-point SFF is known exactly \cite{Saad:2022kfe,okounkov}:
\beq 
\label{eq:KAiryFull}
\begin{split}
K^{(2)}_\text{Airy}(y_1,y_2) &= \frac{e^{S_0 + \frac{\beta^3}{3} \, e^{-2S_0}}}{4\sqrt{\pi}(2\beta)^{3/2}}\, \text{Erf} \left(e^{-S_0}\sqrt{2\beta(\beta^2+T^2)}\right),
\end{split}
\eeq
where we analytically continued $y_{1,2} = \beta \pm i T$. The prefactor of the error function is the asymptotic ``plateau'' $\langle Z_\text{Airy}(2\beta) \rangle \text{ at } T \rightarrow \infty$ (with $\beta$ and $S_0$ fixed).

For the purposes of this Letter we simplify matters by performing the lift of a ``simplified'' model in which we drop the doubly exponential prefactor (the full GUE Airy model will be discussed in \cite{toappear}). This is equivalent to lifting the Airy model in the $\uptau$-scaling limit \cite{Blommaert:2022lbh,Saad:2022kfe} which captures the ramp-plateau transition,
\beq
 T \rightarrow \infty \; \text{ with } \; \uptau \equiv T \, e^{-S_0} \;\text{ fixed}
\eeq
To take this limit, we rescale \eqref{eq:KAiryFull} by $e^{-S_0}$ and observe $\beta^2 + T^2 \sim e^{2S_0}\uptau^2$.

The RMT$_2$ lift of the $\uptau$-scaled GUE Airy model is performed in the two steps prescribed above. First, the Mellin space formulation of the $\uptau$-scaled Airy model is 
\beq
\label{eq:tauAiryf}
\begin{split}
 &f^{(2)}_{\uptau\text{-Airy}}(\omega_1,\omega_2) 
 =\frac{e^{-2i \omega_+S_0}}{6\pi (\omega_++i\epsilon)(2\omega_+ - i)} \\ 
 &\qquad \times
 \Gamma\left(\frac{1}{2}-\frac{i \omega_+}{2}+\frac{3i \omega_-}{2}\right)
 \Gamma\left(\frac{1}{2}-\frac{i \omega_+}{2}-\frac{3i \omega_-}{2}\right) \,, 
\end{split}
\eeq 
where we use the convenient basis of $\omega_\pm = \frac{1}{3}(\omega_1\pm \omega_2)$, and $\epsilon \rightarrow 0$ is a regulator of the pole $\omega_+=0$. This then defines the (Eisenstein sector of the) RMT$_2$ lift, written using $\omega_\pm$:
\begin{equation}
\begin{split}
&{\cal Z}^{(2)}_{\uptau\text{-Airy}}(\{\tau_i\}|{\bf 0}) =\\
 & \frac{9}{8\pi^2}\int d\omega_\pm\,  f^{(2)}_{\uptau\text{-Airy}}(\{\omega_i\})  \left(E_{s_1}(\tau_1) E_{s_2}(\tau_2)\right)_\text{sym}  + \ldots\quad 
 \end{split}
\label{eq:tauAiryRMT2}
\end{equation}
We suppress the zero modes and cusp forms, discussed further in \cite{toappear}.

The topological expansion of the Airy model is encoded in the analytic structure of the overlaps \eqref{eq:tauAiryf}: in particular, poles correspond to fixed-genus contributions. To illustrate this, we imagine performing the inverse Mellin transform to retrieve the SFF $K^{(2)}_{\uptau\text{-Airy}}$, i.e., the integrals \eqref{eq:tauAiryRMT2} with $E_{s_i}(\tau_i) \rightarrow \sqrt{y_i} \,y_i^{i\omega_i}$. 
One can start with the $\omega_-$ integral by closing the contour in either the upper or lower half complex plane. This picks up an infinite series of residues from the $\Gamma$-functions, which resum into 
\beq
\label{eq:KAiryResult}
\begin{split}
&e^{-S_0}\,K^{(2)}_{\uptau\text{-Airy}}(y_1,y_2) = \\
&\;= \frac{1}{4\pi^2}\int_{\mathbb{R}+i\epsilon} d\omega_+ \frac{(y_1y_2)^{\frac{1}{2}+i\omega_+} (y_1+y_2)^{i\omega_+-1}}{e^{(1+2i\omega_+) S_0}} \frac{\Gamma(-i\omega_+)}{(1+2i\omega_+)}
\end{split} 
\eeq
The exponential factor implies convergence of the integral as $\text{Im}(\omega_+) \rightarrow -\infty$; we can therefore perform the $\omega_+$ integral via residues in the lower half complex plane. Relevant poles are located at $\omega_+^{(g)} = -i g$ for $g=0,1,2,\ldots$, and the $g$'th pole is suppressed by $e^{-(2g+1)S_0}$:  one can check that its residue indeed produces the genus $g$ term in the topological expansion of the $\uptau$-scaled Airy model, scaling as $\beta^{g-1}\uptau^{2g+1}$. 

We stress a noteworthy aspect of this expansion: \textit{the higher-genus terms are increasingly ``off-diagonal"}, localized at $\omega_1+\omega_2 = -3ig$. Note that the $g=0$ pole, on the diagonal, precisely yields the double-scaled RMT result \eqref{eq:Kramp}, whose diagonality in Mellin space was previously understood as a CFT${}_2$ avatar of Berry's approximation in periodic orbit theory \cite{berry,DiUbaldo:2023qli}.\\

Having defined the RMT${}_2$ lift \eqref{eq:tauAiryRMT2} of the $\uptau$-scaled Airy model, we can leverage our understanding of its analytic structure to develop the {\it modular-invariant genus expansion}, thus revealing its inherently two-dimensional nature via modular corrections. This expansion takes the form
\beq
\label{eq:ZRMT2Eisensteins}
\begin{split}
&
{\cal Z}^{(2)}_{\uptau\text{-Airy}}(\tau_1,\tau_2|{\bf 0})
= {\cal Z}^{(0,2)}(\tau_1,\tau_2) + \\
&\;\;+ \sum_{g=1}^\infty \frac{(-1)^g\, e^{-2gS_0}}{2\pi g (2g+1)}\sum_{k= 0}^{g-1} \frac{\left(E_{2g-k}(\tau_1) E_{g+k+1}(\tau_2)\right)_\text{sym}}{\Gamma(g-k)k!}\,,
\end{split}
\eeq
where the genus $g=0$ contribution ${\cal Z}^{(0,2)}$ is universal, i.e., identical to (the Eisenstein part of) \eqref{eq:CJRMT2}.

This modular-invariant topological expansion has several interesting features. One of them is the appearance of integer-index Eisenstein series with index bounded above by the genus; in this context, note that replacing $E_n(\tau_i) \rightarrow y_i^{n-1/2}$ in the second line of \eqref{eq:ZRMT2Eisensteins} reproduces exactly the topological expansion of the $\uptau$-scaled Airy RMT \eqref{eq:KAiryFull}. Another feature is the asymptotic character of the sums \eqref{eq:ZRMT2Eisensteins} at late time: a nontrivial resummation ensures that the late-time plateau in the $\uptau$-scaling limit of RMT is preserved in RMT$_2$ despite the modular corrections at every genus $g$ being larger than the original RMT terms at genus $g+1$. The $\uptau$-scaled limit of the Airy RMT \eqref{eq:KAiryFull} is thus recovered,
\beq 
 \lim_{\substack{T \rightarrow \infty \\ \uptau \text{ fixed}}} \, e^{-S_0}\,{\cal Z}^{(2)}_{\uptau\text{-Airy}}(\tau_1,\tau_2|{\bf 0}) = \langle Z_{\text{Airy}}(2\beta)\rangle \, \text{Erf}\left(\uptau \sqrt{2\beta}\right) ,
\eeq
using the analytic continuation $\tau_{1,2} = x_{1,2} + i (\beta \pm i T)$.\\

Finally, we note that the $\uptau$-scaled Airy correlator \eqref{eq:tauAiryf} is paradigmatic for a much larger class of models. In \cite{toappear} we show that for any spectral density $\rho_0(E)$ with a square root edge the $\uptau$-scaled SFF in the GUE ensemble is encoded in 
\beq
\label{eq:fgeneral}
f^{(2)}_{\uptau\text{-RMT}(\rho)}(\omega_1,\omega_2) = f^{(2)}_{\uptau\text{-Airy}}(\omega_1,\omega_2) \times \mathfrak{h}_{\rho}(\omega_+) \,.
\eeq
In particular, the dependence on $\omega_-$ is universal and the way in which the poles encode the genus expansion is structurally identical to the Airy case above.

\section{Application: off-shell wormholes in AdS$_3$ pure gravity}

\label{sec:applications}

By applying this machinery to RMT correlators dual to wormhole amplitudes in two-dimensional gravity, RMT$_2$ makes predictions for fully connected Euclidean wormhole amplitudes with multiple torus boundaries in AdS$_3$ gravity. These are off-shell amplitudes, generalizing \cite{Cotler:2020ugk}, for three-manifolds $\mathcal{M}_3$ of topology $\Sigma_{0,n} \times S^1$ with trivial fibration and boundary topology $\partial \mathcal{M}_3 = \mathbb{T}^2 \cup \cdots \cup \mathbb{T}^2$, the union of $n$ disjoint tori. So far there are no explicit computations or predictions, from either boundary or bulk, for these wormholes with $n>2$, either for AdS$_3$ pure gravity or with matter. \\

\paragraph*{\bf Proposal.}

Our proposal for computing AdS${}_3$ gravity amplitudes for $\Sigma_{0,n} \times S^1$ is simply to plug the appropriate RMT correlators into (\ref{eq:mellin}) and compute (\ref{eq:ZRMT2def}). Let us henceforth focus on AdS${}_3$ pure gravity. To leading order in the semiclassical limit, the proposal is to uplift JT gravity amplitudes, i.e. double-scaled RMT correlators with spectral curve $\rho_0(E) = {\gamma\over 2\pi^2}\sinh(2\pi\sqrt{2\gamma E})$, on topology $\Sigma_{0,n}$:
\beq\label{gravclaim}
 {\cal Z}_{\text{AdS}_3}^{(0,n)}(\{\tau_i\}) \propto {\cal Z}^{(0,n)}_{\text{JT-RMT}_2}(\{\tau_i\}) \,,
\eeq
where we denote $ {\cal Z}_{\text{AdS}_3}^{(0,n)}(\{\tau_i\})$ as the pure gravity amplitude on $\Sigma_{0,n} \times S^1$. 

Moreover, in the scalar sector ${\bf m} = {\bf 0}$, the explicit form of the uplifted JT correlators can be obtained through a simple \textit{replacement rule}, where monomials in the $\{y_i\}$ are replaced by Eisenstein series of appropriate index:
\beq
{\cal Z}^{(0,n)}_{\text{JT-RMT}_2}(\{\tau_i \}|{\bf 0}) = 
K_{\text{JT}}^{(0,n)}(\{y_i\})\Big|_{y_i^{a} \,\mapsto\, E_{{1\over 2}+a}(\tau_i)} 
+ \big[\substack{\text{\tt cusp}\\ \text{\tt forms}}\big]
\label{eq:replacement_rule}
\eeq
with suitable regularization of the $a=1/2$ case. We derive \eqref{eq:replacement_rule} in Appendix \ref{app:n5wormhole}.  The proportionality symbol in \eqref{gravclaim} signals an unspecified overall $\tau_i$-independent normalization of the gravitational path integral.

This prescription, an $n$-point instantiation of the MaxRMT proposal \cite{DiUbaldo:2023qli}, is supported by the known emergence of Schwarzian dynamics in the near-extremal limit of 2D CFTs \cite{Mertens:2017mtv,Ghosh:2019rcj}, the dual emergence of JT dynamics in the dimensional reduction of AdS$_3$ gravity \cite{Maxfield:2020ale}, and the general mechanism of RMT$_2$ presented in this Letter. As noted earlier, RMT$_2$ amplitudes contain the RMT seed amplitudes, via the near-extremal limit of $y\propto \gamma$ and $\gamma\rightarrow \infty$; in the dimensional reduction of pure AdS$_3$ gravity to JT gravity, $\gamma \approx \frac{c}{24}$ at large $c$ \cite{Mertens:2017mtv,Ghosh:2019rcj,Maxfield:2020ale}. Note that while AdS$_3$ gravity also contains off-shell wormholes of fixed boundary topology but with ``interior'' bulk topology, expected to be suppressed by factors exponentially small in $G_N$, the RMT$_2$ prescription above gives the leading-order AdS$_3$ amplitude for a given boundary topology by uplifting the leading-order JT amplitude. 

We now demonstrate the RMT$_2$ prescription for the three-boundary torus wormhole in AdS$_3$ pure gravity. Because of a special universality of the $n=3$ case, we are able to perform a heuristic gravity calculation, which is found to match the RMT$_2$ result. We then describe the RMT${}_2$ lift of JT gravity amplitudes for arbitrary $n$. We record the explicit results for $n=4,5$ in Appendix \ref{app:n5wormhole}.\\

\paragraph*{\bf Three-boundary wormhole from RMT${}_2$.}

We start from the three-point RMT correlator, which is universal for any spectral curve $\rho_0(E)$ with square root edge \cite{Eynard:2015aea}:
\beq
 K^{(0,3)}_\text{RMT}(y_1,y_2,y_3) = \frac{e^{-S_0}}{(2\pi\gamma)^{3/2}} \, \sqrt{y_1y_2y_3}  \,,
\eeq
where the inverse temperatures $y_i$ are measured in units of $\gamma$. The Mellin transform of a monomial naively vanishes (e.g. \cite{zagier}), but admits an $\epsilon$-prescription (e.g. Appendix B of \cite{Penedones:2019tng}). This leads to the spectral overlap
\beq
 f^{(0,3)}_\text{RMT}(\omega_1,\omega_2,\omega_3) = \frac{e^{-S_0}}{(2\pi\gamma)^{3/2}}\prod_{i=1}^{3} \frac{8\epsilon}{4\epsilon^2-(1-2i\omega_i)^2} \,,
\eeq
where $\{\omega_i\}$ are constrained to lie on the contours $\text{Im}(\omega_i) = -1/2$. Note that the overlap manifestly factorizes. We now compute the amplitude ${\cal Z}^{(0,3)}_{\text{RMT}_2}(\tau_1,\tau_2,\tau_3)$ by plugging this overlap into the spectral integral with the contour specified above. On each boundary we must evaluate the integral \eqref{E1int}, which must be regularized, as the Eisenstein series has a pole at $s=1$. Using the principal value prescription $\text{P.V.} \int d\tilde\omega/[(\epsilon^2 + \tilde\omega^2)\tilde\omega] = 0$ 
leads to the final result:
\beq\label{rmt2n3}
 {\cal Z}^{(0,3)}_{\text{RMT}_2}(\{\tau_i\}|{\bf 0}) = \frac{e^{-S_0}}{(2\pi\gamma)^{3/2}}  \widehat{E}_1(\tau_1) \widehat{E}_1(\tau_2) \widehat{E}_1(\tau_3) + \big[\substack{\text{\tt cusp}\\ \text{\tt forms}}\big]
\eeq
where $\widehat{E}_1(\tau)$ is the regular part of the Eisenstein series at $s=1$,
\beq
\label{eq:E1hat}
 \widehat{E}_1(\tau) \equiv \lim_{s\rightarrow 1} \left( E_s(\tau) - \frac{3}{\pi(s-1)}\right)\,.
\eeq

We posit that (up to overall normalization) \eqref{rmt2n3} is the AdS$_3$ pure gravity amplitude for the off-shell three-boundary wormhole $\Sigma_{0,3} \times S^1$, plus terms encoding near-extremal statistics in spinning sectors \`a la \eqref{eq:ZRMT2defgen}; these terms will be suppressed by entropic factors $\sim e^{-\#\sqrt{c}}$. (A natural extension of our result to accommodate spinning sectors, via Poincare sum over an RMT seed in every spin sector, was also considered in \cite{deBoer:2025rct}.)

Let us make a couple of remarks. First, one should view the principal value prescription as a choice of regularization. While this is a natural choice, more generally there is a possibility of adding a $\tau_i$-independent constant on each boundary component; consistently with our holographic claim, this feature is represented on the gravity side as well (see below). Second, note that the result does not admit an expansion into a discrete sum over Virasoro characters on each boundary torus, consistent with a coarse-grained interpretation of semiclassical AdS$_3$ pure gravity in which such amplitudes capture higher moments of dual CFT spectral densities.\\

\paragraph*{\bf Three-boundary wormhole from gravity.}
We now perform a heuristic gravitational computation of the amplitude \eqref{rmt2n3} for the three-boundary wormhole with topology $\Sigma_{0,3}\times S^1$, utilizing certain simplifications that occur for $n=3$. We reiterate that this is not a first principles bulk computation, which we understand is being pursued elsewhere \cite{toappearCCJY}. See also the Supplemental Material for more details.

Let us first illustrate the calculation by analogy with the two-boundary wormhole. After summing the seed amplitude in \eqref{eq:CJresult} over $T$ transformations  $\tau\rightarrow\tau+n$, the resulting seed $g_{\text{AdS}_3}^{(0,2)}(\tau_1,\tau_2)$ can be written as 
\begin{align}
&\frac{g_{\text{AdS}_3}^{(0,2)}(\tau_1,\tau_2)}{\sqrt{y_1y_2}}=2 \frac{\sqrt{y_1y_2}}{|\tau_1\tau_2|}\sum_{m\in \mathbb{Z}} \int_{0}^{\infty}dP_1^2dP_2^2d\bar{P}_1^2d\bar{P}_2^2 \\
&\qquad\qquad\qquad \times\qty|Z_{\rm Tr}(\tau_1,P_1) Z_{\rm Tr}(\tau_2,P_2)|^2 V^{(m)}_{0,2}(\{P_i\})\,\notag,
\end{align}
where $Z_{\rm Tr}(\tau,P)$ is the ``chiral trumpet''  ($S\tau\equiv-1/\tau$)
\beq
Z_{\rm Tr}(\tau,P)\equiv \eta(\tau)\chi_{P}\qty(S\tau), \qquad \chi_{P}(\tau)= \frac{e^{2\pi i\tau P^2}}{\eta(\tau)}
\eeq
\\
which reduces to the well-known two-dimensional trumpet by setting $\tau=i\beta$ \cite{Cotler:2020ugk,Collier:2023cyw,Teschner:2003em,Takhtajan:1993vt,Hadasz:2005gk,Harrison:2022frl,Colville:2023nry}. The trumpets are glued together along geodesic lengths $P_1,P_2$ by a spin-quantized generalization of the Weil-Petersson volumes:
\beq
V^{(m)}_{0,2}=\delta(P_1^2-P_2^2)\delta(\bar{P}_1^2-\bar{P}_2^2)\delta(P_1^2-\bar{P}_1^2-m).
\eeq
The Eisenstein part of the two-boundary wormhole is obtained by considering only the $m=0$ term and summing over $\text{SL}(2,\mathbb{Z})$.
The overall factor is the symplectic volume form on the wormhole moduli space \cite{Cotler:2020ugk}: 
\beq
\sqrt{|\Omega(S\tau_1,S\tau_2)|}=\sqrt{\text{Im}(S\tau_1)\text{Im}(S\tau_2)}=\frac{\sqrt{y_1y_2}}{|\tau_1\tau_2|}\,.
\eeq
This motivates the following gravitational ansatz for the part of the Eisenstein sector of the three-boundary wormhole amplitude that encodes near-extremal RMT statistics in the scalar sector. We introduce
\begin{widetext}
\beq
\begin{split}
\frac{g_{\text{AdS}_3}^{(0,3)}(\{\tau_i\})}{\sqrt{y_1y_2y_3}} &\supset \sqrt{|\Omega(S \tau_1,S \tau_2, S\tau_3)|} \,
  \left| \int_0^\infty dP_1^2dP_2^2dP^2_3 \, Z_\text{Tr}(\tau_1,P_1)Z_\text{Tr}(\tau_2,P_2)Z_\text{Tr}(\tau_3,P_3) V_{0,3}(\{P_i\}) \right|^2  
  \,.
\end{split}
\eeq
\end{widetext}
The object $g_{\text{AdS}_3}^{(0,3)}(\{\tau_i\})$ is defined as the seed of a triple Poincare sum over $\text{SL}(2,\mathbb{Z})/\Gamma_\infty$, one on each boundary. The prefactor is taken to be $\sqrt{|\Omega( \tau_1, \tau_2, \tau_3)|}=\sqrt{y_1y_2y_3}$ in analogy with the two-boundary case. The volume $V_{0,3}(\{P_i\})$ is taken to be the Weil-Petersson volume of a three-holed sphere, namely, $V_{0,3}=1$. In Supplemental Material we discuss this choice further; we only mention here that $V_{0,3}=1$ was obtained explicitly in \cite{Post:2024itb} as a regularized volume within Virasoro TQFT, using a special property of the three-punctured case. The ``$\supset$'' indicates that we are dropping terms coming from the enforcement of RMT statistics in spinning sectors, exponentially small in $\sqrt{c}$, recalling that we have focused throughout this Letter on the ${\bf m}={\bf 0}$ sector of \eqref{eq:ZRMT2defgen}.
 
Combining these ingredients, the above expression yields the simple result $g_{\text{AdS}_3}^{(0,3)}(\{\tau_i\}) \supset y_1y_2y_3/(2\pi)^{6}$. Summing over modular images on each boundary gives the gravity amplitude:
\beq
 {\cal Z}_{\text{AdS}_3}^{(0,3)}(\tau_1,\tau_2,\tau_3) \supset \prod_{i=1}^3 \left( \sum_{\gamma \in \text{SL}(2,\mathbb{Z})/\Gamma_\infty} \text{Im}(\gamma \tau_i) \right) \,,
\eeq 
where we have dropped the overall constant normalization. Each sum is the Poincar\'e series representation of the Eisenstein series $E_{s}(\tau_i)$ at $s=1$. As this is singular, it must be regularized; this is precisely the same freedom of regularization scheme that appears on the RMT${}_2$ side. A minimal scheme is to subtract off the pole. This yields a match with \eqref{rmt2n3}.\\

\paragraph*{\bf $n$-boundary wormholes from RMT${}_2$.}

One can perform the RMT${}_2$ lift of JT gravity amplitudes for arbitrary $n$. For all $n>3$, the amplitude ${\cal Z}^{(0,n)}_{\text{JT-RMT}_2}$ is a polynomial in Eisenstein series, as governed by the replacement rule \eqref{eq:replacement_rule}. As in the $n=3$ case, it does not admit an expansion into a discrete sum over Virasoro characters on the $n$ boundary tori, and reduces to $K^{(0,n)}_\text{JT}$ in the near-extremal limit $y\propto \gamma$ and $\gamma \rightarrow \infty$. The microcanonical spectral statistics of the dual CFT$_2$ follow from these results via inverse Laplace transform. 

We note that for $n>3$, the most naive extension of the heuristic gravity calculation above does not match the RMT${}_2$ result. This is as expected, from considerations of the $n>3$ bulk moduli space.
A bona fide AdS$_3$ pure gravity calculation of the $n$-point amplitudes would be of clear value in learning about random statistics of AdS$_3$ black hole microstates: either AdS$_3$ pure gravity does indeed furnish the minimal completion of random matrix statistics, ratifying the proposal \eqref{gravclaim}-\eqref{eq:replacement_rule}; or, perhaps pure gravity is richer than that, containing additional structure that appears only in higher-point correlations.\\ 

\section{Outlook}
\label{sec:outlook}

In this Letter we presented a prescription that lifts correlation functions in arbitrary random matrix theories to two-dimensional modular-invariant form factors. We presented the GUE Airy model as the simplest example, and used the formalism to predict multi-boundary off-shell wormhole amplitudes in AdS${}_3$ pure gravity. In \cite{toappear} we will discuss more involved examples as well as general spectral curves, topological recursion, other universality classes such as the GOE ensemble, and late-time dynamics. We will discuss the cusp form part of RMT$_2$ and RMT statistics at fixed spin. We will also elaborate on the embedding of RMT${}_2$ into CFT${}_2$, and on the novel constraints imposed by RMT${}_2$ on the spectra of chaotic CFT${}_2$. 

We recall a different approach towards random matrix universality in CFTs: the matrix-tensor model of \cite{Belin:2023efa,Jafferis:2024jkb}, which describes an ensemble of CFT data which {\it approximately} solves the CFT bootstrap constraints up to some tolerance. One awaits fully explicit calculations of correlators in that model; these would enable quantitative comparison to both RMT${}_2$ and semiclassical gravity calculations of off-shell wormhole amplitudes. Our perspective suggests that, upon integrating out the tensor degrees of freedom (which encode the OPE dynamics of the CFT), the resulting spectral matrix model lies within an RMT${}_2$ universality class after taking a suitable limit of vanishing tolerance. This would be very interesting to pursue.

We end with a comment on the broader point of our proposal. Lacking explicit examples of irrational chaotic CFTs, as well as a first-principles gravity calculation for multi-boundary wormholes, we wish to emphasize that RMT${}_2$ predictions should be independently viewed as a way to benchmark chaos in any modular-invariant theory. This rationale is similar to the bootstrap program: RMT${}_2$ encapsulates the constraints that any modular-invariant form factor needs to satisfy if it is to be consistent with random matrix universality. We have argued that this follows from the construction of RMT${}_2$ using a complete basis of modular-invariant functions and minimal universal input near extremality. Holographically, we envision RMT${}_2$ predictions as a baseline against which to compare future gravity calculations; such a comparison would quantify the degree to which AdS${}_3$ gravity is richer than perhaps expected, containing information that goes beyond a reduction to JT gravity near extremality combined with symmetry constraints.

\vspace{1cm}

\begin{acknowledgments}
The authors are grateful to A.\ Blommaert, S.\ Collier, A.\ Etkin, T.\ Hartman, K.\ Jensen, B.\ Post, P.\ Saad, G.\ Wong, C.\ Yan and S.\ Yao for helpful discussions.
The work is supported in part by UK Research and Innovation (UKRI) under the UK government’s Horizon Europe funding Guarantee EP/X030334/1, by Japan Science and Technology Agency (JST) as part of Adopting Sustainable Partnerships for Innovative Research Ecosystem (ASPIRE), Grant No.\ JPMJAP2318, by Institut Pascal at Universit\'{e} Paris-Saclay with the support of the program ``Investissements d’avenir'' ANR-11-IDEX-0003-01, by Grant No.\ NSF PHY-2309135 to the Kavli Institute for Theoretical Physics (KITP), by a Discovery grant from NSERC and by ERC Starting Grant No.\ 853507. J.B.\ and G.D.\  thank the University of British Columbia for hospitality.
\end{acknowledgments}

\appendix

\onecolumngrid


\section{Wormholes with $n>3$ boundaries}
\label{app:n5wormhole}
For wormholes of topology $\Sigma_{0,n} \times S^1$ with $n>3$ boundaries, the RMT$_2$ result is sensitive to the choice of spectral curve of the RMT  being lifted. As motivated earlier, the pure gravity amplitudes on $\Sigma_{0,n} \times S^1$ to leading order in the semiclassical limit should be given by the RMT$_2$ lift of the JT gravity amplitudes on $\Sigma_{0,n}$. We emphasize that the procedure can be repeated for any given $n$, using the genus-0, $n$-boundary JT correlators in \cite{Mertens:2020hbs,Maxfield:2020ale}.

In general, the $n$-boundary JT wormhole amplitudes take the form of a sum of monomials $\prod_i y_i^{a_i}$. The tool that allows us to find the relevant overlaps for each individual term is the regularized Mellin transform introduced in \cite{Penedones:2019tng}:
\begin{equation}
 {\cal M}[y^a](-i\omega) \equiv \int_0^\infty dy \; y^{a-1-i\omega} = \frac{2\epsilon}{\epsilon^2 - (a-i\omega)^2} \,,\qquad
 \text{Im}(\omega) = -a \,.
\end{equation}
The overlaps $f^{(0,n)}_{\text{JT}} (\{ \omega_i \})$ are sum of products of the above expression. Because of this factorized structure, the modular uplift acts individually on each monomial and reduces to integrals of the form
\beq\label{Eisint}
\begin{split}
\lim_{\epsilon \to 0}
\left( 
\int_{\mathbb{R}- i a} 
\frac{d\omega}{2\pi} 
\frac{2\epsilon}{\epsilon^2-(a-i \omega)^2}  \,
E_{\frac{1}{2}+i \omega} (\tau) 
\right)  
 = E_{\frac{1}{2}+a} (\tau) \qquad\qquad &\Big(a\neq {1\over 2}\Big)\,.
 \end{split}
 \eeq
The case $a=1/2$ is slightly more subtle because of the Eisenstein pole at $s=1$: the right-hand side of \eqref{Eisint} should be understood as the pole-subtracted Eisenstein series $\widehat{E}_1(\tau)$, defined in \eqref{eq:E1hat}. To arrive at this result, we expand the integrand near $\omega=0$ and use the integral representation of the Dirac delta function to obtain 
\beq\label{E1int}
\lim_{\epsilon\rightarrow 0} \int_{\mathbb{R}} \frac{d\omega}{2\pi}  \frac{2\epsilon}{\epsilon^2 +\omega^2} \, E_{1+i\omega} (\tau) = \widehat{E}_1(\tau) + \lim_{\epsilon\rightarrow 0}\int_{\mathbb{R}} \frac{d\omega}{2\pi}  \frac{2\epsilon}{\epsilon^2 +\omega^2} \frac{3}{\pi i \omega} \,.
\eeq
The second term, a $\tau$-independent constant, must be regularized: a canonical choice is the principal value prescription,
\beq\label{PV}
\text{P.V.} \left[\int_{\mathbb{R}} {d\omega\over (\epsilon^2 + \omega^2)\omega}\right] = 0\,.
\eeq
Adopting this choice henceforth, this allows us to immediately write down the modular uplift of JT gravity amplitudes -- indeed, of {\it any} polynomial RMT amplitude -- by using the replacement rule
\begin{equation}
   y^a \; \longmapsto \; E_{\frac{1}{2}+a}(\tau) - \frac{3}{\pi(a-\frac{1}{2})} \, \delta_{\frac{1}{2},a}\,.
   \label{eq:replacement}
\end{equation}
This proves \eqref{eq:replacement_rule} and also applies to the all-genus expansion of the $\uptau$-scaled Airy model, see \eqref{eq:ZRMT2Eisensteins}.

For further illustration, we now present the RMT$_2$ prediction for $n=4$ and  $n=5$ wormhole amplitudes in AdS${}_3$ pure gravity by lifting corresponding JT gravity expressions.\\

\textbf{$n=4$ boundaries.} The JT gravity amplitude with four boundaries is
\beq 
K^{(0,4)}_{\rm JT}(\{y_i \}) 
= 
\frac{e^{-2S_0}}{4\pi^2 \gamma^3} \sqrt{y_1 y_2 y_3 y_4}\left(2\pi^2 \gamma+\sum_{i=1}^4 y_i \right)\,,
\label{eq:JT_n=4_answer}
\eeq
where $\gamma = c/24$ when JT gravity is embedded as a near-extremal sector of AdS$_3$ pure gravity. Using the replacement rule \eqref{eq:replacement} leads to the following four-boundary RMT$_2$ amplitude: 
\begin{align}
\label{eq:RMT2_n=4_answer}
& {\cal Z}^{(0,4)}_{\text{JT-RMT}_2}
 (\{ \tau_i \}|{\bf 0}) 
= \frac{e^{-2S_0}}{4\pi^2 \gamma^3} \left( \prod_{i=1}^4 
 \widehat{E}_{1} (\tau_i)\right) 
\left( 
2\pi^2 \gamma 
+ 
\sum_{i=1}^4 \frac{E_2(\tau_i) }{\widehat{E}_1 (\tau_i)} 
\right) + [\text{\tt cusp forms}]
\,
.
\end{align}
Importantly for the consistency of our proposal,
in the near-extremal limit  $y_i \propto \gamma$ and $\gamma \to \infty$, one recovers the JT gravity $n=4$ wormhole \eqref{eq:JT_n=4_answer}.  Note that using instead the VMS spectral curve for $n=4$ amounts simply to having $\gamma_{\rm VMS}=\frac{c-13}{24}$ in the above expressions, with the same near-extremal limit reproducing $K^{(0,4)}_\text{VMS}(\{y_i\})$. \\

\textbf{$n=5$ boundaries.}
The JT gravity amplitude for a wormhole with $n=5$ boundaries and trivial interior topology is given by 
\beq
\label{eq:KJT5}
K^{(0,5)}_\text{JT}(\{y_i \})  = \frac{e^{-3S_0}}{4\sqrt{2}\pi^{5/2}\gamma^{9/2}}\, \sqrt{y_1y_2y_3y_4 y_5} \left[10\pi^4 \gamma^2+6\pi^2 \gamma\left(\sum_{i=1}^5 y_i\right) +  \left(\sum_{i=1}^5 y_i\right)^2 \right]
\eeq
The RMT$_2$ prediction for the five-boundary amplitude follows by expanding \eqref{eq:KJT5} and applying \eqref{eq:replacement} term by term:
\begin{equation}
\label{eq:RMT2_n=5_answer}
\begin{split}
& {\cal Z}^{(0,5)}_{\text{JT-RMT}_2}
 (\{ \tau_i \}|{\bf 0}) 
\\&\quad =
\frac{e^{-3S_0}}{4\sqrt{2}\pi^{5/2}\gamma^{9/2}}
\left( \prod_{i=1}^{5} \widehat{E}_{1}(\tau_i)  \right) \left[ 
10 \pi^4 \gamma^2 +  \sum_{i=1}^5 \left(6\pi^2 \gamma\,
\frac{E_2(\tau_i)}{\widehat{E}_1 (\tau_i)} 
+
\frac{E_3(\tau_i) }{\widehat{E}_1(\tau_i)} \right) +
2\sum_{i<j} 
\frac{E_2(\tau_i)E_2(\tau_j)}{\widehat{E}_1 (\tau_i)\widehat{E}_1 (\tau_j)}
\right]+ \big[\substack{\text{\tt cusp}\\ \text{\tt forms}}\big]~.
\end{split}
\end{equation}

For arbitrary $n$, $K_\text{JT}^{(0,n)}(\{ y_i \})$ is only known as a generating function, but one may proceed in the above fashion for any given $n$.

\section{Details on gravity computation of three-boundary wormhole amplitude}
 
In this appendix we motivate and examine further our gravitational computation of the off-shell three-boundary wormhole amplitude in AdS$_3$ pure gravity, of topology $\Sigma_{0,3} \times S^1$. We emphasize that this is a heuristic computation, not rigorously derived from the gravitational path integral. Nevertheless, in somewhat the same spirit as the hallmark computation of Cotler and Jensen of the two-boundary wormhole \cite{Cotler:2020ugk}, there are good reasons to expect that it may indeed be correct, thanks to degenerate features of the three-boundary case. 

We recall our expression for convenience: up to an overall constant normalization, the gravity amplitude is
\beq
\mathcal{Z}_{{\rm AdS}_3}^{(0,3)}(\tau_1,\tau_2,\tau_3) \supset \left(\prod_{i=1}^3 \sum_{\gamma_i \in \text{SL}(2,\mathbb{Z})/\Gamma_\infty}\right)g_{\text{AdS}_3}^{(0,3)}(\gamma_1\tau_1,\gamma_2\tau_2,\gamma_3\tau_3)
\eeq
with seed amplitude
\beq\label{eq:gravansatz}
\begin{split}
\frac{g_{\text{AdS}_3}^{(0,3)}(\{\tau_i\})}{\sqrt{y_1y_2y_3}} &\supset \sqrt{|\Omega(S \tau_1,S \tau_2, S\tau_3)|} \,
  \left| \int_0^\infty dP_1^2dP_2^2dP^2_3 \, Z_\text{Tr}(\tau_1,P_1)Z_\text{Tr}(\tau_2,P_2)Z_\text{Tr}(\tau_3,P_3) V^{(0)}_{0,3}(\{P_i\}) \right|^2 =\frac{\sqrt{y_1y_2y_3}}{(2\pi)^{6}}.
\end{split}
\eeq
The ''$\supset$'' indicates that, in the language of our general RMT$_2$ framework \eqref{eq:ZRMT2defgen}, this is the ${\bf m}={\bf 0}$ piece of the gravity amplitude, which encodes the RMT statistics near extremality in the scalar sector on all three boundaries. Note that spinning terms are suppressed by at least one factor of $e^{-S_0(j)}$, where $S_0(j) \approx 2\pi \sqrt{cj/ 6}$, yielding exponential suppression in the semiclassical limit $c = {3/(2G_N)} \rightarrow\infty$ . 

This result uses multiple structural assumptions:
\begin{enumerate}[label=(\arabic*)]
\item Holomorphic factorization, up to the measure factor $\sqrt{|\Omega(S \tau_1,S \tau_2, S\tau_3)|}$. 

\item The choice of measure for integrating over moduli: $\sqrt{|\Omega(S \tau_1,S \tau_2, S\tau_3)|} = \prod_{i=1}^3 \sqrt{\text{Im}(S\tau_i)} = \prod_{i=1}^3  {\sqrt{y_i}/ |\tau_i|}$.  

\item Identification of the volume: $V^{(0)}_{0,3}(\{P_i\}) = 1$.

\item The choice of measure for gluing the trumpets: $\prod_{i=1}^3 dP_i^2$.

\end{enumerate}

\noindent We now proceed to explain and analyze these assumptions.  

\vskip .1 in

 The expression is holomorphically factorized, up to the overall measure factor. This is naturally the case when considering {\it hyperbolic} (on-shell) three-manifolds in AdS$_3$ pure gravity \cite{Collier:2023fwi}. The phase space on hyperbolic manifolds of the form $\Sigma_{g,n}\times \mathbb{R}$ is locally holomorphically factorized and given by two copies $\mathcal{T}\times \bar{\mathcal{T}}$ of the Teichm{\"u}ller component of the moduli space of flat $\text{SL}(2,\mathbb{R})$ bundles on $\Sigma_{g,n}$.\footnote{Teichm{\"u}ller space is the universal cover of moduli space and thus only gives a local description of the initial value surface $\Sigma_{g,n}$,  not accounting for global identifications under the mapping class group ${\rm MCG}(\Sigma_{g,n})$.} 
 The  Hilbert space is obtained by the quantization of Teichm{\"u}ller space and consequently it is holomorphically factorized \cite{Collier:2023fwi,EllegaardAndersen:2011vps,andersen2013new,Chekhov:1999tn,kashaev1998quantization,VERLINDE1990652,Teschner:2005bz,Teschner:2003at,Teschner:2003em}. This is the Hilbert space of Virasoro topological quantum field theory, or Virasoro TQFT (VTQFT). 
 As a result, the partition functions of hyperbolic wormholes with topology $\Sigma_{g,n}\times I$ are holomorphically factorized.\footnote{To compute the 3d gravity path integral on a hyperbolic manifold $\mathcal{M}_3$ one needs to further sum over geometries which corresponds to summing over elements of the quotient $\rm{MCG}(\partial \mathcal{M}_3)/\rm{MCG}(\partial \mathcal{M}_3, \mathcal{M}_3)$. Only individual terms in the sum are holomorphically factorized.}
 
The case of wormholes of topology $\Sigma_{0,n}\times S^1$ is more subtle: they are {\it non-hyperbolic} (off-shell) geometries. In particular, VTQFT is not, in its original definition, able to correctly capture non-hyperbolic amplitudes, as we will further discuss below.\\

To frame our discussion, let us recall the salient properties of the $n=2$ wormhole of topology $\mathbb{T}^2 \times I$ \cite{Cotler:2020ugk}. There is a nontrivial integration measure $\Omega$ over the moduli space of wormhole geometries which accounts for the mapping class group. The moduli space consists of physical parameters characterizing the wormhole geometry, namely the minimal lengths in given homotopy classes and the corresponding holonomies and relative twists. If instead one keeps such moduli {\it fixed} in the gravitational path integral, it is possible to find moduli-dependent on-shell saddles for the previously off-shell geometry. These saddles are called constrained instantons \cite{Affleck:1980mp,Cotler:2020lxj,Cotler:2021cqa} since their on-shell actions  explicitly depend on the moduli and are not minimized with respect to them.\footnote{An example in 2d dilaton gravity theories, whereupon fixing the length of the wormhole it becomes a solution, is discussed in Appendix A of \cite{Stanford:2020wkf}.} To complete the path integral one must integrate over the moduli space with measure given by the symplectic volume form $\sqrt{|\Omega|}\prod_i d\mu(m_i)$, where $d\mu(m_i)$ schematically denotes the measure over moduli. The moduli couple to the boundary metric, so the factor $\sqrt{|\Omega|}$  explicitly depends on $\tau_i$ and spoils holomorphic factorization. Before integration however, the fixed-moduli wormholes are holomorphically factorized. More precisely, the path integral over the fixed moduli wormholes is given by two pairs (one for each boundary) of holomorphic and antiholomorphic Alekseev-Shatashvili theories of boundary reparametrizations $\rm{Diff}(S^1)/U(1)$, resulting in two pairs of trumpets $|Z_{\rm Tr}(\tau_1,P)|^2$ and $|Z_{\rm Tr}(\tau_2,P)|^2$ \cite{Cotler:2018zff,Cotler:2020ugk}. The integral over moduli glues the trumpets together, properly accounting for the mapping class group $\rm{MCG}(\Sigma_{0,2})$ via the  Weil-Petersson volume of moduli space $V_{0,2}(P_1,P_2)$. After summing over relative $T$ transformations, one obtains the spin-quantized Weil-Petersson volume $V^{(j)}_{0,2}(\{P_i\})$ As for the measure $\Omega$, this was argued in \cite{Cotler:2020ugk} to be $|\Omega| = y_1y_2/|\tau_1\tau_2|^2$. That choice is crucial to achieve near-extremal RMT behaviour, e.g. the linear ramp of the SFF. \\
  
Returning now to the $n=3$ case of the three-boundary wormhole, the ansatz \eqref{eq:gravansatz} assumes a similar structure:  holomorphically factorized trumpets with fixed moduli, and a symplectic volume form dictating how to integrate over the moduli space. In other words, our gravitational ansatz \eqref{eq:gravansatz} can be understood as adopting the following perspective: the $\Sigma_{0,3} \times S^1$ wormhole amplitude is equal to the holomorphically factorized product that one would compute for a genuine saddle point geometry, augmented by the overall factor $\sqrt{|\Omega|}$ that reinstates the non-hyperbolicity. We justify this further below. In addition, we have made a particular choice of symplectic measure which mimics the choice in the $n=2$ case; we stress that it needs to be justified from a first principles treatment of the gravity path integral, carefully accounting for the mapping class group.

Within this form of ansatz, we now address the treatment of the ``chiral trumpets'', which we have glued via the Weil-Petersson volume $V^{(0)}_{0,3}=1$ and measure $\prod_{i=1}^3 dP_i^2$. We discuss these, and the overall form of the ansatz, from two separate perspectives: the Virasoro TQFT and the Virasoro Minimal String. \\

\textbf{Virasoro TQFT.}
The volume $V^{(0)}_{0,3}$ can be independently motivated from a recent explicit (albeit regularized) computation in VTQFT. As recalled above, the applicability of VTQFT \cite{Collier:2023fwi} is currently limited to hyperbolic three-manifolds, as it neglects the quotient by ${\rm{MCG}}(\Sigma_{g,n})$ necessary to describe the moduli space $\mathcal{M}(\Sigma_{g,n})= \mathcal{T}(\Sigma_{g,n})/\rm{MCG}(\Sigma_{g,n})$.\footnote{See the recent paper \cite{Yan:2025usw} for discussion on the application of VTQFT to non-hyperbolic geometries leading to incorrect 3d gravity results.} Given these limitations it seems that VTQFT has little to tell us about the three-boundary wormhole. However, for $n=3$, this is not the case: for the three-punctured sphere, $\rm{MCG}(\Sigma_{0,3})$ is trivial. This property does not hold for $n>3$.
The authors of \cite{Post:2024itb} explicitly obtained the volume $V^{(0)}_{0,3}=1$ from VTQFT by considering a manifold $\mathcal{M}_3^O$ with the topology of the three-boundary wormhole, augmented with a Wilson line stretching between two of the boundaries, which is then taken to vanish. The insertion of a Wilson line of fixed conformal weight $h_O$ renders the manifold hyperbolic. The limit $h_O\rightarrow 0$, where the Wilson line is taken to the identity, is convergent and results in
 \beq
V_{0,3}(\{P_i\}) \equiv \lim_{h_O\rightarrow 0} Z_{\rm Vir}(h_O;\{P_i\})=1.
 \eeq
The identification of $V_{0,3}(\{P_i\})$ with this limiting quantity is to be understood as a sort of regularization, albeit a convergent one.

The fact that this VTQFT volume computation gives a finite result that is relevant for 3d gravity {\it per se} should be viewed as rather special to $n=3$ and not indicative of the general case: as we have repeatedly remarked, for $n>3$ the mapping class group is non-trivial. We also note that, were one to attempt  an analogous computation nevertheless at higher (or lower) $n$, limits where a Wilson line is taken to the identity are typically divergent. For example, the analogous computation in the $n=2$ case \cite{Chandra:2022bqq,Yan:2023rjh,Yan:2025usw} has a $1/h_O$ pole, whose residue is not the empty two-boundary wormhole of Cotler and Jensen but rather a different wormhole amplitude \cite{Collier:2023fwi,Cotler:2020ugk,Cotler:2020hgz} which does not contain any RMT behaviour. This issue is also present in JT gravity \cite{Yan:2023rjh}.

This brings us to the issue of gluing measure for the trumpets, namely, our choice of measure flat in $P_i^2$. If one uses VTQFT to glue asymptotic trumpets to $V_{0,3}$ in a holomorphically factorized way, this is not expected to produce the 3d gravity result \eqref{eq:gravansatz}, even though the theory computes a finite $V_{0,3}$: the measure for gluing trumpets in VTQFT is $\prod_i dP_i$, not $\prod_i dP_i^2$. Whereas the latter is the one necessary to generate random matrix dynamics, the former is what the inner product between Virasoro blocks generates. The origin of the discrepancy is known: the VTQFT computation does not take into account the relative twist when gluing geodesics. Integrating over the twist gives the measure $\prod_i dP_i^2$, as is well known in 2d \cite{Saad:2019lba}. For this reason we adopt the measure flat in $P_i^2$, again highlighting the need to derive this independently from gravity.
\\

\textbf{Virasoro Minimal String.}
 In \cite{Collier:2023cyw} it was shown that quantizing a ``chiral half of 3d gravity" on manifolds $\Sigma_{g,n}\times S^1$ with geodesic boundaries $P_{1},\dots, P_n$ results in the ``quantum volumes" of the VMS: denoting the chiral partition function as $Z_{\rm VMS}^{\chi}$, 
 \beq
Z_{\rm VMS}^{\chi}(\Sigma_{g,n}\times S^1;\{P_i\})=V_{g,n}^{(b)}(\{P_i\})\,,
 \eeq
where $b$ is related to a CFT central charge ($c=1+6(b+b^{-1})$) and appears in the (Cardy) spectral curve,
\begin{equation}
    \rho_{0,\text{VMS}}(E) \propto \frac{2\sqrt{2}}{ \sqrt{E}}\,\sinh\left(2\pi b \sqrt{E}\right) \sinh\left(2\pi b^{-1} \sqrt{E}\right) \,.
\end{equation}
This is a one-parameter generalization of the JT spectral curve, and accordingly, the VMS volumes $V_{g,n}^{(b)}$ are generalizations of the Weil-Petersson volumes of JT gravity. The problem of quantizing 3d gravity on a fixed geometry is simplified by considering only a chiral half of the theory, and subsequently gluing an anti-chiral half. In return, within each half, one obtains the ability to correctly quotient by the mapping class group (in contrast to VTQFT). As a result we obtain the volumes of moduli space for $\Sigma_{g,n}$ with arbitrary $g,n$.

Our gravitational ansatz \eqref{eq:gravansatz} can therefore be understood as adopting the perspective that the $\Sigma_{0,3} \times S^1$ wormhole amplitude admits a description as a gluing of chiral and anti-chiral gravity amplitudes, augmented by a moduli measure factor that is not reproduced by either half. We have motivated above why this is reasonable for the $n=3$ case specifically. Upon adopting this point of view, the quantization of the (anti-)chiral half of 3d gravity then leads directly to the use of $V_{0,3}=1$ in the ansatz. Moreover, this approach automatically generates the trumpet gluing measure $\prod_i dP^2_i$, which follows from the Weil-Petersson measure used to compute the volumes $V^{(b)}_{g,n}$.

\twocolumngrid 

\bibliographystyle{apsrev}


\end{document}